\documentclass{article}
\usepackage{spconf,amsmath,graphicx,hyperref}
\usepackage{cite}
\usepackage{amsmath,amssymb,amsfonts}
\usepackage{algorithm}
\usepackage{algpseudocode}
\usepackage{bm}
\usepackage{setspace}
\usepackage{graphicx}
\usepackage{subfigure}
\usepackage{subcaption}

\makeatletter
\renewcommand{\fnum@algorithm}{}

\title{Hierarchical Sparse Vector Transmission for Ultra Reliable and Low Latency Communications}
\name{Yanfeng Zhang$^{1}$, Xi'an Fan$^{2}$, Jinkai Zheng$^{1}$, Xiaoye Jing$^{1}$, Weiwei Yang$^{1}$, and Xu Zhu$^{3}$ %\thanks{Thanks to XYZ agency for funding.}
\vspace{-10pt}}
\address{ $^{1}$School of Electrical Engineering \& Intelligentization, Dongguan University of Technology, China\\
$^{2}$School of Computer Science and Technology, Dongguan University of Technology, China\\
$^{3}$School of Electronic and Information Engineering, Harbin Institute of Technology, Shenzhen, China\\\vspace{-20pt}}

\begin{document}
%\ninept
%
\maketitle

\begin{abstract}
Sparse vector transmission (SVT) is a promising candidate technology for achieving ultra-reliable low-latency communication (URLLC). In this paper, a hierarchical SVT scheme is proposed for multi-user URLLC scenarios. The hierarchical SVT scheme partitions the transmitted bits into common and private parts. The common information is conveyed by the indices of non-zero sections in a sparse vector, while each user’s private information is embedded into non-zero blocks with specific block lengths. At the receiver, the common bits are first recovered from the detected non-zero sections, followed by user-specific private bits decoding based on the corresponding non-zero block indices. Simulation results show the proposed scheme outperforms state-of-the-art SVT schemes in terms of block error rate.

%Sparse vector coding (SVC) is a promising short-packet transmission method for ultra reliable low latency communication (URLLC) in next generation communication systems. In this paper, a message-splitting SVC (MS-SVC) based short packet transmission scheme is proposed to further enhance the transmission performance of SVC. The core idea behind the proposed scheme lies in mapping the transmitted information bits onto sparse vectors via block and single-element sparse mappings. The block sparse mapping pattern is able to concentrate the transmit power in a small number of non-zero blocks thus improving the decoding accuracy, while the single-element sparse mapping pattern ensures that the code length does not increase dramatically with the number of transmitted information bits. At the receiver, a two-stage decoding algorithm is proposed to sequentially identify non-zero block indexes and single-element non-zero indexes. Extensive simulation results verify that the proposed MS-SVC scheme outperforms the existing SVC schemes in terms of block error rate and spectral efficiency.
\end{abstract}
\section{Introduction}
\label{sec:intro}

Ultra-reliable low-latency communication (URLLC) is a key service category in next-generation mobile networks, tailored for mission-critical applications like autonomous driving, Internet of Things (IoT), and remote control systems \cite{Kim20201,Autonomous24,IOTURLLC2022}. To meet the stringent low latency requirements of URLLC, short-packet transmission has attracted significant attention from both academia and industry. As a novel short-packet transmission scheme, the sparse vector coding (SVC)  has been widely studied due to its flexible operation and excellent block error rate (BLER) performance in short block length regimes \cite{Ji2018,Kim20202,ZhangXuewan2022,YangLinjie2024,CRushTIT2021,
Sinha2024TCOM,DhanTWC2023,yfzhang2024BSVC,XZHANG2025,Hsieh2021}. SVC maps information bits onto sparse vectors, which are then projected through random spreading codebook and can asymptotically achieve capacity over complex additive white Gaussian noise (AWGN) channels \cite{Hsieh2021}.

%In SVC, the information bits are first mapped onto a sparse vector, which is then projected onto the transmitted signal space through a random spreading codebook \cite{Ji2018}. Its error performance has been demonstrated in \cite{Hsieh2021} to be asymptotically capacity-achieving over the complex additive white Gaussian noise (AWGN) channel.

Various extensions have been developed to improve the performance of original SVC. The enhanced SVC (ESVC) \cite{Kim20202} jointly maps information bits onto non-zero indices and quadrature amplitude modulation (QAM) symbols, while sparse superimposed codes (SSC) \cite{ZhangXuewan2022} exploit multiple constellations to enhance reliability. The constellation indices have been leveraged as auxiliary carriers of information to reduce BLER \cite{YangLinjie2024}. The sparse regression codes (SPARC) \cite{CRushTIT2021} is another important SVC approach, with approximate message passing adopted as a low-complexity decoder. Recent advances include generalized SPARC (GSPARC) \cite{Sinha2024TCOM} and block orthogonal sparse superposition (BOSS) codes \cite{DhanTWC2023}, where the latter constructs codewords via sequential mapping across orthogonal sub-codebooks. Our earlier work proposed a block SVC (BSVC) scheme by introducing block-sparse mapping patterns \cite{yfzhang2025BSVC}. A constellation-label-aided index redefinition (IR) scheme has been introduced in \cite{XZHANG2025} by employing base conversion and index collision shifting to achieve near-optimal compressibility. Furthermore, the concept of SVC has been extended to diverse communication systems, including multiple-input multiple-output (MIMO) systems \cite{ZhangRuoyu2021}, low-resolution ADCs systems \cite{ZYF2024ICCC}, high-mobility communications \cite{ZhangYf2023}, semantic communications \cite{Zhanxunyang25}, grant-free access \cite{Luoyingzhe24}, low storage overhead systems \cite{ZYFWCNC25}, etc.

%In summary, the existing SVC studies mainly focus on the design of sparse mapping patterns and decoding algorithms \cite{Kim20202,ZhangXuewan2022,YangLinjie2024,CRushTIT2021,
%Sinha2024TCOM,DhanTWC2023,yfzhang2024BSVC,XZHANG2025,Hsieh2021}, without considering the transmission latency overhead caused by redundant information in multi-user scenarios. More importantly, the existing multi-user SVC schemes rely on codebook division \cite{ZhangRuoyu2021} or channel characteristics \cite{RSMASVC24} to mitigate inter-user interference, which makes it difficult to ensure high reliability when the number of users increases and the channel varies rapidly.

Existing SVC studies primarily focus on sparse mapping and decoding design \cite{Kim20202,ZhangXuewan2022,YangLinjie2024,CRushTIT2021,
Sinha2024TCOM,DhanTWC2023,yfzhang2024BSVC,XZHANG2025,Hsieh2021}, but overlook transmission latency overhead from redundant information in multi-user scenarios. Moreover, current multi-user SVC schemes rely on codebook division \cite{ZhangRuoyu2021} or channel characteristics \cite{RSMASVC24} to mitigate interference, making it challenging to maintain high reliability as user numbers increase or channels vary rapidly.

In this paper, we propose a hierarchical SVC (HSVC) scheme that partitions multiple users' information bits into common and private components. In the first layer, common information bits are mapped onto the indices of $U$ non-zero sections in a sparse vector, while in the second layer, each user's private information is embedded into non-zero blocks with specific block lengths. We design a successive decoding algorithm that first recovers common information by identifying the $U$ non-zero section indices, then decodes private information based on pre-allocated non-zero block lengths. The proposed HSVC scheme maps all users' common and private information onto the same sparse vector, avoiding redundant transmission of common information, and thus reducing transmission latency. Furthermore, by allocating non-zero blocks with different block lengths to users, the proposed HSVC scheme can mitigate inter-user interference and improve decoding reliability.

%The proposed HSVC scheme maps all users' common information onto the same sparse vector, reducing transmission latency. Furthermore, by assigning different non-zero block lengths to users, the proposed scheme mitigates inter-user interference and enhances decoding reliability.

\section{Proposed HSVC Scheme}
In this section, we introduce the sparse mapping pattern and encoding process of the HSVC scheme.
\vspace{-10pt}
\subsection{HSVC Encoding}
We consider a downlink communication system where a base station (BS) serves $U$ users, as shown in Fig. 1. The BS uses a message splitting pattern to transmit different messages to $U$ users simultaneously.
The transmitted information bits $b$ is partitioned into $U+1$ bit streams, \emph{i.e.}, one common bit stream (traffic lights, surveillance equipments, street lights information, etc.) and $U$ private bit streams (speed up/slow down, turn left/right, and start/stop commands, etc.). The total transmitted bits can be expressed as
\begin{equation}
b = {b_{\rm{c}}} + \sum\nolimits_{u = 1}^U {{b_u}},
\end{equation}
where $b_{\rm c}$ and ${b_u}$ denote the number of common bits and the number of private bits of the $u$-th user, respectively.

As illustrated in Fig. 1, the $b_{\rm c}$ common bits are mapped onto the indices of $U$ non-zero sections in a sparse vector. Specifically, a sparse vector $\bf s$ of length $N$ is uniformly divided into $S$ sections with each of length $D=N/S$, where $U$ ($U<S$) non-zero sections are selected and allocated to $U$ users for transmitting private information. Therefore, the number of common bits can be calculated as
\begin{equation}
{b_{\rm{c}}} = \left\lfloor {{{\log }_2}\left( {\begin{array}{*{20}{c}}
S\\
U
\end{array}} \right)} \right\rfloor,
\end{equation}
where $\left\lfloor  \cdot  \right\rfloor $ denote the round-down operation. For example, given the number of sections $S=4$ and the number of non-zero sections $U=2$, then ${b_c} = \left\lfloor {{{\log }_2}\left( {\scriptstyle4\atop
\scriptstyle2} \right)} \right\rfloor  = 2$ bits of common information can be encoded.

For the $u$-th user, $b_{ u}$ private bits are divided into two parts, where $b_{u,1}$ bits are mapped to $K_u$ non-zero block indices in the $u$-th non-zero section after block sparse mapping, while the remaining $b_{u,2}$ bits are mapped to the non-zero values after QAM modulation. The block sparse mapping for the $u$-th user involves selecting $K_u$ non-zero blocks of length $L_u$ from $D$ positions, thus the number of bits that can be mapped onto the non-zero block indices is
\begin{equation}
{b_{u,1}} = \left\lfloor {{{\log }_2}\left( {\begin{array}{*{20}{c}}
{D - {K_u}({L_u} - 1)}\\
{{K_u}}
\end{array}} \right)} \right\rfloor ,
\end{equation}
where $L_u$ denotes the length of non-zero blocks at the $u$-th non-zero section. The number of bits mapped to QAM symbols by the $u$-th user is given as
\begin{equation}
{b_{u,2}} = {L_u}{K_u}{\log _2}({M_{\bmod }}),
\end{equation}
where ${M_{\bmod }}$ is the order of QAM modulation. For instance, if 4-QAM modulation is adopted, $b_{1,1}+b_{1,2}= \left\lfloor {{{\log }_2}\left( {\scriptstyle9-3\atop
\scriptstyle1} \right)} \right\rfloor  +4\times 1\times 2=10$ private bits are mapped to a non-zero block of length 4 in the first non-zero section (the first user), and $b_{2,1}+b_{2,2}= \left\lfloor {{{\log }_2}\left( {\scriptstyle9-1\atop
\scriptstyle1} \right)} \right\rfloor +2\times 1\times 2 =7$ private bits can be mapped to a non-zero block of length 2 in the second non-zero section (the second user).

\emph{Remark:} The BS allocates non-zero blocks with distinct block lengths to different users, allowing each user to recover its private information using the respective block length information at the receiver.

\begin{figure}[htbp]
\centerline{\includegraphics[width=0.45\textwidth]{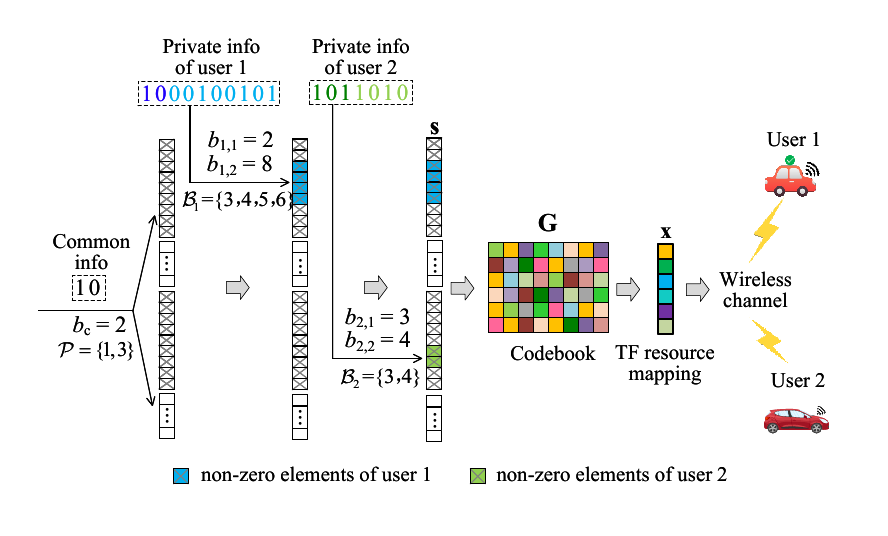}}
\caption{Diagram of HSVC encoding process with $N=36$, $U=2$, $S=4$, $D=9$, $L_{1}=4$ and $L_{2}=2$.}
\label{fig1}
\end{figure}
\vspace{-15pt}

\vspace{-10pt}
\subsection{Random Spreading of HSVC}
After sparse mapping, the non-zero elements in the sparse vector $\bf s$ are spread through pseudo-random codewords. This process can be described as
\begin{equation}
{\bf{x}} = \sum\nolimits_{u = 1}^U {\sum\nolimits_{k = 1}^{{K_u}} {{{\bf{g}}_{u,k}}{{{s}}_{u,k}}} }  , 
\end{equation}
where ${\bf{x}} \in {\mathbb{C}^M}$ denotes the transmitted TF-domain signal, ${{\bf{g}}_{u,k}} \in {{\mathbb R}^{M \times {L_u}}}$ and ${{{{s}}_{u,k}}}$ are the codeword and modulated QAM symbol corresponding to the $k$-th non-zero block of the $u$-th user, respectively. The complete codebook matrix can be expressed as ${\bf{G}} = [{{\bf{G}}_1},{{\bf{G}}_2}, \cdots ,{{\bf{G}}_S}] \in {\mathbb{R}^{M \times N}}$,  where ${{\bf{G}}_i} = [{{\bf{g}}_{i,1}},{{\bf{g}}_{i,2}}, \cdots ,{{\bf{g}}_{i,D}}] \in {\mathbb{R}^{M \times D}}$ is the $i$-th sub-codebook. The elements of codebook are composed of 1 and -1, both following a Bernoulli distribution \cite{Ji2018}. An example of codebook for $S=3$, $D=2$, $M=3$ and $N=6$ is given by
\begin{equation}
{\bf{G}} = \sqrt {\frac{1}{{{K_{{\rm{non}}}}}}} \underbrace {\left[ {\begin{array}{*{20}{r}}
1&{ - 1}\\
{ - 1}&1\\
1&{ - 1}
\end{array}} \right.}_{{{\bf{G}}_1}}\underbrace {{\rm{ }}\begin{array}{*{20}{r}}
{ - 1}&{ - 1}\\
1&{ - 1}\\
1&1
\end{array}}_{{{\bf{G}}_2}}\underbrace {\left. {{\rm{ }}\begin{array}{*{20}{r}}
1&{ - 1}\\
1&1\\
{ - 1}&{ - 1}
\end{array}} \right]}_{{{\bf{G}}_3}},
\end{equation}
where ${K_{{\rm{non}}}} = \sum\nolimits_{u = 1}^U { {{{K_u}L_u}} } $ is the total number of non-zero elements. The resulting spread sequence $\bf x$ is then mapped to time-frequency (TF) resources for transmission.

%\subsection{Transmission of Spread Sequence}
The spread sequence $\bf x$ is converted to time domain after performing inverse discrete Fourier transform (IDFT) on it, \emph{i.e.}, ${{\bf{x}}_{\rm{T}}} = {{\bf{F}}^{\rm H}}{\bf{x}}$, where ${\bf{F}} \in {\mathbb{C}^{{M} \times {M}}}$ is a normalized DFT matrix with ${{\bf F}_{m,n}} = \frac{1}{{\sqrt M }}\exp ( - j2\pi mn/M)$. A $L_{\rm{CP}}$-length cyclic prefix (CP) is added to $\bf{x}_{\rm{T}}$ to avoid inter-symbol interference. At the receiver, after implementation of DFT on received signal and CP removal, the received signal of the $u$-th user can be expressed as
\begin{equation}
\begin{aligned}
{{\bf{y}}_u} &= {\bf{FH}}_{\rm{T}}^u{{\bf{F}}^{\rm{H}}}{\bf{Gs}} + {\bf{w}}\\
 &= {{\bf{\Phi }}_u}{\bf{s}} + {\bf{w}}
\end{aligned},
\label{eq4}
\end{equation}
where ${{\bf{\Phi }}_u} = {\bf{FH}}_{\rm{T}}^u{{\bf{F}}^{\rm{H}}}{\bf{G}} \in {\mathbb{C}^{M \times N}}$, ${{\bf{H}}_{\rm{T}}^u} \in {\mathbb{C}^{M \times M}}$ and ${{\bf{H}}_{\rm{F}}^u} \in {\mathbb{C}^{M \times M}}$ denote the time-domain and frequency-domain channel between the RSU and $u$-th user, respectively. ${\bf{w}} \sim {\cal C}{\cal N}(0,{\sigma ^2}{{\bf{I}}_{M}})$ is the AWGN vector with variance $\sigma ^2$.

%In static or low-speed moving scenarios, the time-domain channel gains remain constant within the duration of a transmission block. Therefore, the frequency-domain channel matrix ${{\bf{H}}_{\rm{F}}^{u}}$ is a diagonal matrix.

%\begin{figure}[htbp]
%\centerline{\includegraphics[width=0.48\textwidth]{figs/MSBSRCmapping.pdf}}
%\caption{Diagram of the sparse mapping process of DM-SVC. $b$ bits of information are mapped to a sparse vector of length $N$ containing $K_{\rm b}$ non-zero blocks and $K_{\rm s}$ non-zero elements.\vspace{0pt}}
%\label{fig2}
%\end{figure}

\section{HSVC Decoding}
In this section, the decoding algorithms for common information and private information are described.
\subsection{Common Information Decoding}

%Note that the SE of the conventional SSC scheme \cite{ZhangXuewan2022} is given as
%\begin{equation}
%{\rm{S}}{{\rm{E}}^{{\rm{ESVC}}}} = \frac{{{b_{\rm{I}}} + {b_{\rm{S}}}}}{{{M^{{\rm{SVC}}}}}},
%\end{equation}
%where ${{b_{\rm{I}}}}={\rm{log}_2}{\tbinom{N}{{{K_{\rm{b}}L+K_{\rm s}}}}} $ denotes the number of bits encoded into the non-zero indexes in the SSC scheme \cite{ZhangXuewan2022}.

%By comparing (10) and (11), it can be observed that the number of bits encoded by the MD-SVC scheme is less than that of the SSC scheme \cite{ZhangXuewan2022} because ${b_{{\rm{I,1}}}} + {b_{{\rm{I,2}}}} < {b_{\rm{I}}}$. However, the SSC scheme requires more subcarriers to ensure the transmission reliability as compared to the DM-SVC scheme. Fig. 2 shows the comparison between SSC scheme \cite{ZhangXuewan2022} and DM-SVC scheme in terms of SE. The total number of non-zero elements in both schemes is set to 6. It can be observed that the SE of the DM-SVC scheme with different parameters is higher than that of the SSC scheme. The DM-SVC scheme achieves more than 12\% and 14\% SE improvement compared to the SSC scheme for QPSK and 16-QAM modulation, respectively. When the parameter $({{K_{\rm{b}}}},L,{K_{\rm{s}}}) =(1, 5,1)$ is adopted, the DM-SVC achieves a two-fold SE improvement compared to the SSC scheme for 16-QAM modulation.

%At the receiver, the $ u $-th user knows the number of non-zero sections $ U $ and the number of non-zero blocks $ K_u $ in the $ u $-th non-zero section ${\bf s}_u$. 
At the receiver, the $ u $-th user needs to first recover the common information from the received signal $ \mathbf{y}_{u} $, i.e., identify the indices of $ U $ non-zero sections from $S$ sections. This process can be described as:
\begin{equation}
{{\cal P}^*} = \mathop {\arg \min }\limits_{|{\cal P}| = U} ||{{\bf{y}}_u} - {{\bf{\Phi }}_u}{{\bf{s}}_{\cal P}}||_2^2,
\end{equation}
where ${{\cal P}^*}$ is the index set of the identified non-zero sections. Considering $\bf s$ is a block $U$-sparse signal, the indices of non-zero sections can be identified by the standard block orthogonal matching pursuit (BOMP) algorithm \cite{YEldar2010}. The BOMP algorithm begins by initializing the residual as ${{\bf{r}}_0} = {{\bf{y}}_u}$. At the $u$-th iteration ($u \ge 1$), we choose the section that is best matched to ${{\bf{r}}_{u - 1}}$, according to
\begin{equation}
{i_u} = \mathop {\arg \min }\limits_i ||{\bf{\Phi }}_{u,i}^H{{\bf{r}}_{u - 1}}||_2^2,
\end{equation}
where ${{\bf{\Phi }}_{u,i}} \in {\mathbb{C}^{M \times D}}$ is the $i$-th sub-matrix of ${{\bf{\Phi }}_u}$. Once the index $i_u$ is choosen, we find ${\bf s}_{u,i}$ as the solution to
\begin{equation}
\mathop {\arg \min }\limits_{{{\{ {{{\bf{\tilde s}}}_{u,i}}\} }_{i \in {\cal P}}}} ||{{\bf{y}}_u} - \sum\limits_{i \in {\cal P}} {{{\bf{\Phi }}_{u,i}}{{{\bf{\tilde s}}}_{u,i}}} ||_2^2,
\end{equation}
where $\cal P$ is the set of chosen indexes $i_j$, $1\le j \le u$. The residual is then updated as
\begin{equation}
{{\bf{r}}_u} = {{\bf{y}}_u} - \sum\limits_{i \in {\cal P}} {{{\bf{\Phi }}_{u,i}}{{\bf{s}}_{u,i}}} .
\end{equation}
After $U$ iterations, the $u$-th user can obtain the indexes of $U$ non-zero sections, i.e., ${\cal P}^*= {\cal P}$, and recover $b_c$ bits of common information. For example, if ${\cal{P}} = \{ 1,3 \}$, the decoded common information bits are `10'.

Note that when $M=D$ and the columns of ${\bf \Phi}_{u,i}$ being orthonormal for each $u$ (the elements across different sub-matrices do not have to be orthonormal), the algorithm does not perform a least-squares minimization over the blocks that have already been selected, but directly updates the residual according to ${{\bf{r}}_u} = {{\bf{r}}_{u - 1}} - {{\bf{\Phi }}_{u,{i_u}}}{\bf{\Phi }}_{u,{i_u}}^H{{\bf{r}}_{u - 1}}$, thereby reducing decoding complexity.

\subsection{Private Information Decoding}

%At the receiver, the decoding process of private information consists of two stages. First, the $u$-th user needs to identify the $u$-th non-zero section based on the assigned non-zero block length. Second, it detects the non-zero block index to decode its private information. Since the power of each non-zero section is proportional to the length of non-zero blocks, sections with larger block lengths exhibit higher transmission power. This power ordering enables the $u$-th user to employ a successive interference cancellation (SIC) strategy for decoding. Specifically, the $u$-th user first detects the index of the non-zero block with the largest block length (highest power), then subtracts the corresponding signal component from the received signal. Subsequently, it detects the index of the non-zero block with the second-largest block length, and continues this iterative process until it successfully detects its own corresponding non-zero block.

At the receiver, private information decoding involves two stages: (1) identifying the $u$-th user's non-zero section based on the assigned block length, and (2) detecting the indices and values of non-zero blocks. Since transmission power is proportional to block length, the $u$-th user employs successive interference cancellation (SIC), iteratively detecting and removing signals from highest to lowest power blocks until reaching its own block.

After identifying the index ${\cal P}_u^*$ of the $u$-th non-zero section, the $ u $-th user needs to detect the indices of $ K_u $ non-zero blocks and estimate the values of $ K_u L_u $ non-zero elements. This problem can be formulated as
\begin{equation}
\{{\cal B}_u^*, {{\bf{\hat c}}_{{{\cal B}_u}}}\} = \mathop {\arg \min }\limits_{|{{\cal B}_u}| = {K_u}} ||{{\bf{y}}_u} - {{\bf{\Psi }}_u}{{\bf{c}}_{{{\cal B}_u}}}||_2^2,
\end{equation}
where ${{\bf{\Psi }}_u} = {[{{\bf{\Phi }}_u}]_{{\cal P}_u^*}} \in {{\mathbb{R}}^{M \times D}}$, ${\bf{c}} = {{\bf{s}}_{{\cal P}_u^*}} \in {\mathbb{C}^D}$, and ${\cal B}_u^*$ represents the index set of $K_u$ non-zero blocks corresponding to the $u$-th user. 

%The problem formulated in (12) can be solved using the maximum likelihood (ML) method. To find the ML solution, we need to enumerate all possible candidate index combinations, with cardinalities given by ${\log _2}\left({\scriptstyle {D - {K_u}({L_u} - 1)} \atop \scriptstyle K_u} \right)$. Unfortunately, the exhaustive search method would not be feasible in most practical situations when $D$ and $K_u$ take large values. 
The existing MMP algorithm \cite{Kwon2014} provides a solution to the problem in (12). Nevertheless, the MMP algorithm neglecting the inherent structured sparsity characteristics in $\bf c$, degrading its decoding performance. Moreover, the $K_u$ non-zero blocks in the $u$-th non-zero section are non-uniformly distributed, the conventional BOMP algorithm \cite{YEldar2010} cannot be directly used to solve (12). 

To address this issue, a multi-path BOMP (MBOMP) algorithm is proposed to decoding the private information. According to the circular shift property of matrix-vector operations, one can obtain 
\begin{equation}
{{\bf{y}}_u} = {{\bf{\Psi }}_u}{\bf{c}} + {\bf{w}}{\rm{ = }}{{\bf{\Psi }}_u}{{\bf{\Pi }}^l}{{\bf{\Pi }}^{ - l}}{\bf{c}} + {\bf{w}},
\end{equation}
where ${\bf{\Pi }} = {\mathop{\rm circ}\nolimits} \{ [0,1,0, \cdots ,0]_{M \times 1}^{\rm{T}}\} $ is a permutation matrix. Eq. (13) reveals that the original equation remains invariant when both the sparse vector $\bf{c}$ and the columns of measurement matrix undergo identical circular shifts of $l$ positions. Exploiting this property, we can cyclically shift the vector 
$\bf{c}$ up to $L-1$ times to ensure that at least one non-zero block aligns with the uniformly distributed block positions. 

The MBOMP algorithm first performs $L$ cyclic shifts on the columns of ${\bf {\Psi}}_u$, and utilizes these new measurement matrices to construct $L$ optimization problems:
\begin{equation}
\left\{ {\begin{array}{*{20}{c}}
{{{\hat {\cal B}}_{u,1}} = \mathop {\arg \min }\limits_{|{{\cal B}_u}| = {K_u}} ||{{\bf{y}}_u} - {{\bf{\Psi }}_u}{{\bf{\Pi }}^0}{{\bf{c}}_{{{\cal B}_u}}}||_2^2}\\
{{{\hat {\cal B}}_{u,2}} = \mathop {\arg \min }\limits_{|{{\cal B}_u}| = {K_u}} ||{{\bf{y}}_u} - {{\bf{\Psi }}_u}{{\bf{\Pi }}^1}{{\bf{c}}_{{{\cal B}_u}}}||_2^2}\\
 \vdots \\
{{{\hat {\cal B}}_{u,L }} = \mathop {\arg \min }\limits_{|{{\cal B}_u}| = {K_u}} ||{{\bf{y}}_u} - {{\bf{\Psi }}_u}{{\bf{\Pi }}^{L - 1}}{{\bf{c}}_{{{\cal B}_u}}}||_2^2}
\end{array}} \right..
\end{equation}
Then, the standard BOMP algorithm is applied to solve the $L$ problems in Eq. (14), yielding $L$ candidate support sets ${\cal I} = \{ {{\hat {\cal B}}_{u,1}},{{\hat {\cal B}}_{u,2}}, \cdots ,{{\hat {\cal B}}_{u,L}}\} $. The support set that minimizes the residual is selected as the optimal solution, i.e.,
\begin{equation}
{\cal B}_u^* = \mathop {\arg \min }\limits_{{{\cal B}_u} \subset {\cal I}} ||{{\bf{y}}_u} - {[{{\bf{\Psi }}_u}]_{{{\cal B}_u}}}{{\bf{c}}_{{{\cal B}_u}}}||_2^2,
\end{equation}
the non-zero values of the $u$-th user can be estimated as
\begin{equation}
{{{\bf{\hat c}}}_{{\cal B}_u^*}} = {([{{\bf{\Psi }}_u}]_{{\cal B}_u^*}^H{[{{\bf{\Psi }}_u}]_{{\cal B}_u^*}})^{ - 1}}[{{\bf{\Psi }}_u}]_{{\cal B}_u^*}^H{{\bf{y}}_u}.
\end{equation}
Finally, $b_{u,1}$ bits of information and $b_{u,2}$ bits of information can be recovered through block sparse demapping and soft symbol demodulation \cite{yfzhang2025BSVC}, respectively.

\vspace{-10pt}
\section{Simulation Results}
In this section, the BLER performance of HSVC scheme over Rayleigh fading channels is evaluated. The SVC \cite{Ji2018}, ESVC \cite{Kim20202}, IR-SSC \cite{XZHANG2025}, BOSS \cite{DhanTWC2023} and GSPARC \cite{Sinha2024TCOM} schemes are selected for performance comparison. It is assumed that perfect channel is available in decoding for all schemes. The Rayleigh channel model is adopted. The average BLER is defined as the mean ratio of the number of erroneously received packets to the total number of transmitted packets for all users. In Fig. 2 (a) and Fig. 3 (a), the parameters of HSVC scheme are set to $N=130$, $D=2$, $S=65$, $L_1=2$, $L_2=1$ and $b=15$ bits. In Fig. 2 (b), the parameters of HSVC scheme are set to $N=1032$, $D=12$, $S=86$, $L_1=6$, $L_2=4$, $L_3=3$, $L_4=2$ and $b=35$ bits. All comparison schemes maintain the same number of subcarriers $M$ and total transmitted bits. The comparison schemes adopt a sequential transmission mode where common information is transmitted at first, followed by private information.

\begin{figure}[H]
    \centering
    \subfigure[]{
        \includegraphics[width=0.18\textwidth]{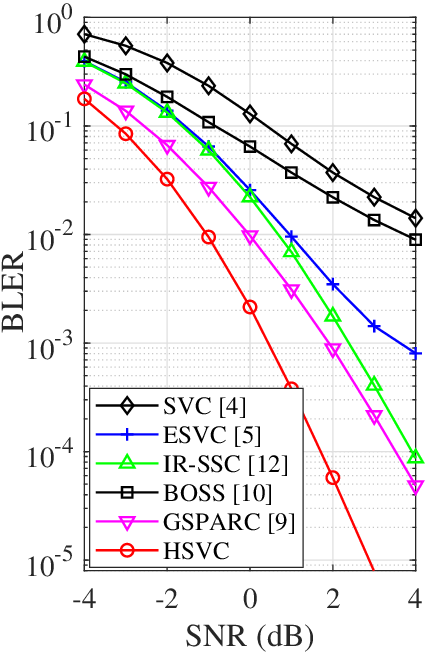}
        \label{fig:sub1}
    }
    \hspace{-8pt}
    \subfigure[]{
        \includegraphics[width=0.18\textwidth]{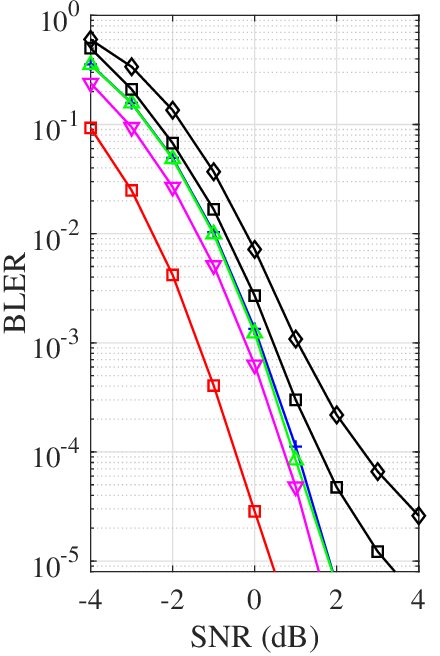}
        \label{fig:sub2}
    }\vspace{-10pt}
    \caption{BLER performance of different schemes for (a) 2 users, and (b) 4 users.}
    \label{fig:main}
\end{figure}
\vspace{-10pt}

\begin{figure}[H]
    \centering
    \subfigure[]{
        \includegraphics[width=0.18\textwidth]{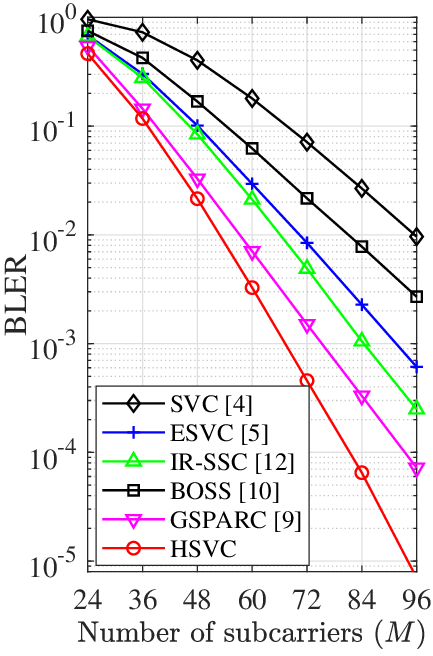}
        \label{fig:sub1}
    }
    \hspace{-8pt}
    \subfigure[]{
        \includegraphics[width=0.18\textwidth]{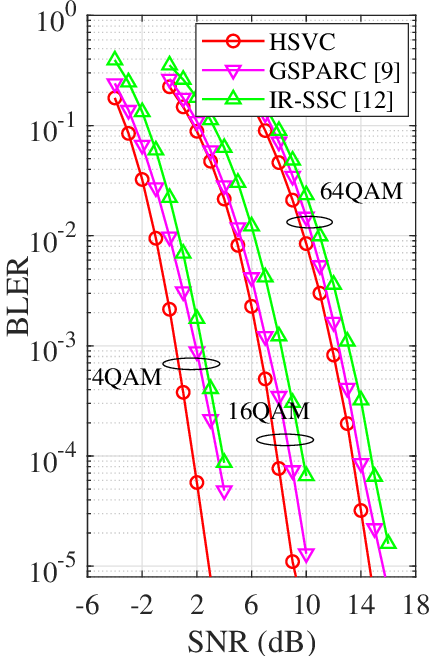}
        \label{fig:sub2}
    }\vspace{-10pt}
    \caption{BLER performance: (a) as a function of subcarrier number $M$ with $\text{SNR}=2$ dB, and (b) comparison of different schemes for various modulation orders.}
    \label{fig:main}
\end{figure}
\vspace{-10pt}

Fig. 2 shows the BLER performance of different schemes, where the proposed HSVC scheme outperforms its competitors. For example, at the target BLER = $10^{-4}$, proposed HSVC achieves SNR gains of 1.8 dB and 1.5 dB over the GSPARC \cite{Sinha2024TCOM} for 2-user and 4-user cases, respectively. This is primarily because existing SVC schemes transmit common and private information sequentially, while the proposed HSVC scheme maps all users' common information onto the same sparse vector and transmits it simultaneously to all users, thus avoiding redundant transmission of common information. Furthermore, the block mapping pattern and block-sparsity-based decoding algorithm employed in HSVC further enhance the BLER performance.

Fig. 3(a) presents the BLER performance as a function of the number of transmitted subcarriers. It can be observed that at BLER = $10^{-4}$, the HSVC scheme achieves a 12\% reduction in subcarrier number compared to the GSPARC scheme \cite{Sinha2024TCOM}. This indicates that proposed HSVC scheme can effectively reduce transmission latency by simultaneously transmitting common and private information. Figure 3(b) shows the BLER performance under various modulation orders, where the HSVC scheme shows no error floor and maintains performance advantages over existing GSPARC \cite{Sinha2024TCOM} and IR-SSC \cite{XZHANG2025} schemes even under high-order modulation.

\section{Conclusions}
In this paper, the HSVC scheme is proposed for short-packet URLLC. The proposed HSVC scheme maps multiple users' common and private information onto a sparse vector and transmits it simultaneously to all users, avoiding redundant transmission of common information and thus reducing transmission latency. Thanks to the block-sparse mapping patterns and decoding algorithms, the HSVC scheme achieves superior BLER performance compared to state-of-the-art SVC schemes. Future work will focus on extending HSVC to multi-user massive MIMO scenarios.

\vfill\pagebreak
\clearpage


\bibliographystyle{IEEEbib}
\begin{spacing}{0.5} 
\bibliography{refs}
\end{spacing}

\end{document}